# Hybrid Template Canonical Correlation Analysis Method for Enhancing SSVEP Recognition under data-limited Condition

Runfeng Miao, Li Zhang, and Qiang Sun

*Abstract*— In this study, an advanced CCA-based algorithm called hybrid template canonical correlation analysis (HTCCA) was proposed to improve the performance of brain-computer interface (BCI）based on steady state visual evoked potential (SSVEP) under data-limited condition. The HTCCA method combines the training data from several subjects to construct SSVEP templates. The experimental results evaluated on two public benchmark datasets showed that the proposed method outperforms the compared methods in both detection accuracy and information transfer rate when the number of trials is small. Considering that user-friendly experience will become a key factor for BCI in practical application, it is very necessary to develop effective methods based on limited EEG samples. This study demonstrates that the proposed method has great potential in the application of SSVEP-based brain-computer interfaces.

## I. INTRODUCTION

Brain computer interface (BCI) based on steady state visual evoked potential (SSVEP) has been widely concerned owing to its less user training, higher information transmission rate (ITR) and higher signal-to-noise ratio (SNR) [1]. SSVEP is a neural response evoked at the same frequency of the visual stimulus and the corresponding harmonic frequencies [2].

Recently, many algorithms have been proposed to detect SSVEPs. Nakanish *et al.* [3] applied task-related component analysis (TRCA) to SSVEP, greatly improving the performance of SSVEP-based BCIs. Wei *et al.* [4] proposed training data-driven canonical correlation analysis (TDCCA) that only uses training data to construct spatial filters, and they proved the equivalence of TDCCA and TRCA theoretically. Yuan *et al.* [5] proposed transfer template based CCA (tt-CCA) method, adding stimulus-related information from the training data of other subjects into the template to enhance SSVEP detection. All these extension methods used calibration data recorded from subjects in a training stage for SSVEP recognition. Subjects may suffer from visual fatigue during a long experiment, which results in performance degradation [6][7]. However, reducing the number of trials will have a significant negative impact on detection performance [3][7]. To get better performance and reduce user fatigue, it is necessary to develop an algorithm which can achieve better performance when the number of trials is small.

In this study, we proposed a novel method, hybrid template canonical correlation analysis (HTCCA), which uses the training information of the specific subject as well as other subjects to construct templates and spatial filters for enhancing target detection, considering that EEG data from each individual subject are limited, while EEG data from many other subjects are available. To prove the effectiveness of the proposed method, we compared the proposed method with the TDCCA method and the tt-CCA method based on two publicly available SSVEP datasets.

## II. MATERIALS AND METHODS

### A. Transfer template-based canonical correlation analysis (tt-CCA)

In the rest of study, $N_c$ denotes the number of channels, $N_s$ indicates the number of sampling points, $N_h$ is the number of harmonics, $N_t$ represents the number of trials and $N_f$ represents the number of total targets.

The tt-CCA method uses two templates. One is the sine-cosine template $Y_i \in R^{2N_h \times N_s}$ which contains a series of sine and cosine signals with specific frequencies as fundamental components and their harmonics and the other is the transferred EEG template $\overline{X_i} \in R^{N_c \times N_s}$. The transferred template $\overline{X_i}$ is created by averaging EEG data across other subjects. There are two spatial filters used in tt-CCA, namely $w_x$ generated by using CCA on the testing data $x \in R^{N_c \times N_s}$ and the pre-constructed template $Y_i$, where $T$ denotes matrix transpose, and $w_{\overline{x}}$ generated by using CCA on the transferred EEG template $\overline{X_i}$ and the template $Y_i$. The Pearson correlation coefficient $\rho_1(f_i)$ is computed between $w_{\overline{x}}^T \overline{X_i}$ and $w_x^T x$, the Pearson correlation coefficient $\rho_2(f_i)$ is computed between $w_x^T \overline{X}$ and $w_x^T x$, and the canonical correlation coefficient $\rho_3(f_i)$ is calculated between $x$ and $Y_i$. At last, the stimulus target is detected according to the maximal sum of those three correlations. The identification formula is as follows,

$$f_{\text{target}} = \arg\max_{f_i}(\rho_1(f_i) + \rho_2(f_i) + \rho_3(f_i)), i = 1, 2, ..., N_f \quad (1)$$

*Research supported by the National Natural Science Foundation of China (Project No. 51977020).

Runfeng Miao is with Chongqing University - University of Cincinnati Joint Co-op Insitute, China (e-mail: miaorg@ mail.uc.edu).

Li Zhang is with the State Key Laboratory of Power Transmission Equipment & System Security and New Technology, School of Electrical Engineering, Chongqing University Electrical Engineering Department, China.

Qiang Sun is with the State Key Laboratory of Power Transmission Equipment & System Security and New Technology, School of Electrical Engineering, Chongqing University, China.

## B. Training data-driven canonical correlation analysis (TDCCA)

Assume that $X_i^n \in R^{N_c \times N_s}$ is the training data of $n$-th trial from the specific subject corresponding to the $i$-th stimulus, and $x \in R^{N_c \times N_s}$ is the testing data. The average-trail matrix, $S_i = \overline{X_i} \in R^{N_c \times N_s}$, is obtained by averaging the training data across all trials. Then, the synthetic matrix is obtained by concatenating every single-trial training data, $E_i = [X_i^1, X_i^2, ..., X_i^{N_t}] \in R^{N_c \times (N_s \cdot N_t)}$. The template is obtained by concatenating the repeated average-trial templates: $Y_i = [S_i, S_i, ..., S_i] \in R^{N_c \times (N_s \cdot N_t)}$.

The spatial filter $w_i$ is generated by using the standard CCA method on $E_i$ and $Y_i$. Then, the Pearson correlation coefficient $\rho(f_i)$ is calculated between $w_i^T x$ and $w_i^T \overline{X_i}$. Finally, the detection result is decided as follows,

$$f_{\text{target}} = \arg\max_{f_i} \rho(f_i), i = 1, 2, ..., N_f \qquad (2)$$

## C. Hybrid template canonical correlation analysis (HTCCA)

Fig. 1 shows the block diagram of the proposed method. For the $i$-th stimulation frequency, assume $X_i^n \in R^{N_c \times N_s}$ is the training data of $n$-th trial of the specific subject, $\overline{X_{ti}^n} \in R^{N_c \times N_s}$ is the training data obtained by averaging all $n$-th trials across subjects except the specific subject., and $x \in R^{N_c \times N_s}$ is the testing data of the specific subject.

In the first stage, $S_i = \overline{X_i} \in R^{N_c \times N_s}$ is obtained by averaging the specific training data across $N_t$ trails, where $N_t$ represents the number of specific training trials. $K_i \in R^{N_c \times N_s}$ is obtained by averaging the training trials from other subjects across $N_t'$ trials, where $N_t'$ indicates the number of independent training trials. The specific template is obtained by concatenating average-trial template $S_i$, $Y_i = [S_i, S_i, ..., S_i] \in R^{N_c \times (N_s \cdot N_t)}$. In the same way, concatenating $K_i$ can yield the independent template, $Y_i' = [K_i, K_i, ..., K_i] \in R^{N_c \times (N_s \cdot N_t')}$. The specific EEG signal is obtained by concatenating all single-trial specific training data, $E_i = [X_i^1, X_i^2, ..., X_i^{N_t}] \in R^{N_c \times (N_s \cdot N_t)}$. The independent EEG signal is acquired by concatenating all single-trial independent training data $\overline{E_{ti}} = [\overline{X_{ti}^1}, \overline{X_{ti}^2}, ..., \overline{X_{ti}^{N_t'}}] \in R^{N_c \times (N_s \cdot N_t')}$. Then, $w_i$ is obtained by using CCA on $E_i$ and $Y_i$, and $w_{ti}$ is obtained by using CCA on $\overline{E_{ti}}$ and $Y_i'$.

In the second stage, the two spatial filters $w_i$ and $w_{ti}$ are used to transform signals as $w_i^T x$, $w_{ti}^T x$, $w_i^T S_i$, $w_i^T K_i$, $w_{ti}^T S_i$ and $w_{ti}^T K_i$, where $x$ represents the testing data.

In the last stage, the Pearson correlation coefficient $\rho_1(f_i)$ is computed between $w_i^T x$ and $w_i^T S_i$, the Pearson correlation coefficient $\rho_2(f_i)$ is computed between $w_i^T x$ and $w_i^T K_i$, the Pearson correlation coefficient $\rho_3(f_i)$ is calculated between $w_{ti}^T x$ and $w_{ti}^T S_i$, and the Pearson correlation $\rho_4(f_i)$ is computed between $w_{ti}^T x$ and $w_{ti}^T K_i$. A weighted correlation coefficient $\rho(f_i)$ is defined as follows,

$$\rho(f_i) = \sum_{n=1}^{4} \text{sign}(\rho_n(f_i)) \rho_n(f_i)^2 \qquad (3)$$

where sign() is sign function, which is the key function to maintain knowledge from negative correlation coefficients. Finally, the target frequency corresponding to the testing data is decided as follows,

$$f_{\text{target}} = \arg\max_{f_i} \rho(f_i), i = 1, 2, ... N_f \qquad (4)$$

## D. Benchmark data and performance evaluation

The first dataset used in this study is Tsinghua dataset [8]. The EEG data were recorded from 35 healthy subjects (seventeen females, mean age was 22 years). 40 stimuli from 8 Hz to 15.8 Hz with an interval of 0.2 Hz were displayed on the screen. For each subject, the data contained six blocks and each block contained 40 trials corresponding to all 40 stimuli. Each trial lasted 6 seconds, which consisted of 0.5 s for visual cues and 5 s for stimulation, then the screen was blank for 0.5 s before next trial. To avoid user fatigue, a rest was set between two blocks. The latency delay time was considered as 140 ms.

The second dataset is San Diego database provided in [9]. EEG data were collected from 10 healthy subjects with 12 targets (nine males, mean age was 28 years). Each trial consists of 15 blocks. In every single block, the subject was asked to look at one of the visual stimuli which was randomly appeared for 4 s. There were 12 trials in each block corresponding to all 12 targets. The BioSemi ActiveTwo EEG system (BioSemi, Inc.) was used to record the EEG data at a sampling rate of 2048 Hz. The recorded EEG data were down-sampled to 256 Hz. The latency delay time was considered as 135ms.

The leave-one-out cross-validation strategy is used to evaluate the detection accuracy of those four methods. For the standard CCA method, since no training process is required, the identification accuracy is evaluated directly through six verifications. In addition to the detection accuracy, the BCI performance is also evaluated by ITR (bits/min) [3][7][9]:

$$ITR = \frac{60}{T} \left\{ \log_2 N + P \log_2 P + (1+P) \log_2 \frac{(1-P)}{(N-1)} \right\} \qquad (5)$$

where $P$ represents the identification accuracy. $T = T_w + T_s$. The data length is represented by $T_w$ and $T_s$ is the time for subjects diverting their attention between two consecutive trials. $N$ indicates the number of targets.

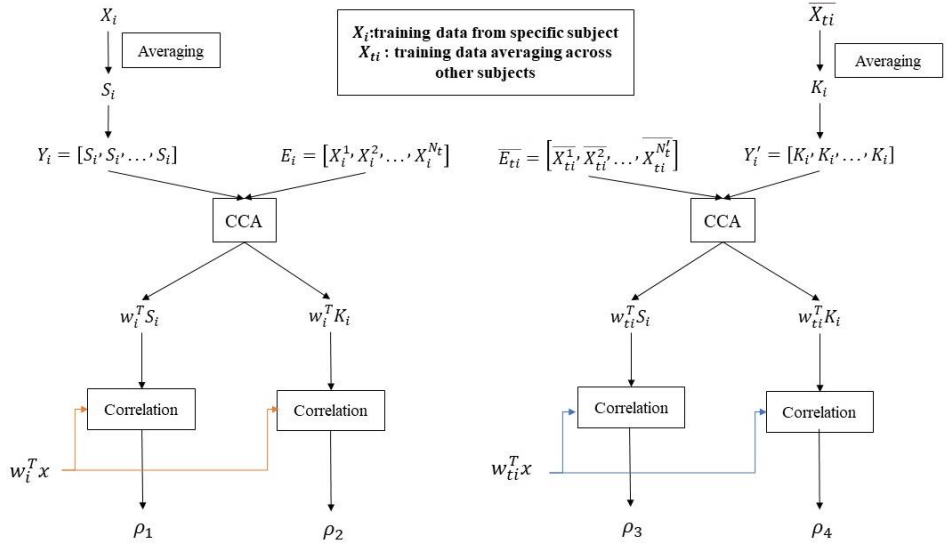

Fig. 1. Block diagram of proposed method.

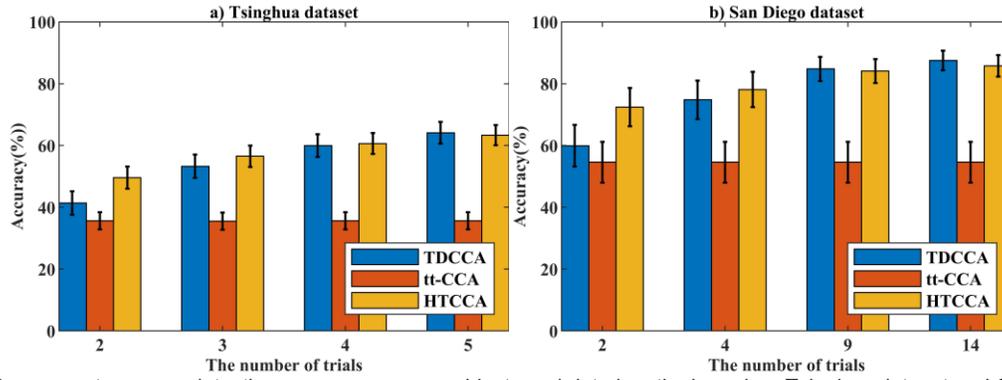

Fig. 2. (a) and (b) represent average detection accuracy across subjects and data lengths based on Tsinghua dataset and San Diego dataset, respectively. Error bars denote the standard error.

## III. RESULTS

For Tsinghua dataset, the results were evaluated by fixing the number of electrodes to nine. And for San Diego dataset, the number of selected electrodes was eight.

Fig.2 shows that the classification performance averaged across subjects and data lengths with different numbers of training trials. Here, to meet the requirement of TDCCA and HTCCA methods (averaging training data is necessary), the minimal number of training trials should be two. Specifically, for San Diego dataset, the number of trials was selected as 2,4,9, and 14, respectively. It can be observed that TDCCA and HTCCA outperformed tt-CCA. HTCCA is slightly inferior to TDCCA when the number of trials is big enough. When the trials were limited, HTCCA achieved superior performance. Paired t-tests showed that there were significant differences between HTCCA and TDCCA when the number of trials equals to 2 ($p<0.001$) and 3 ($p<0.05$) on Tsinghua dataset and when the number of trials equals to 2 ($p<0.001$) and 4 ($p<0.05$) on San Diego dataset, and TDCCA and HTCCA methods were significantly better than tt-CCA method ($p<0.001$ in all conditions, paired t-test). In the following analysis, we will focus only on these two approaches.

Fig.3 shows the detection performance and ITR across subjects with different data lengths when the number of training trials equals to 2 on both Tsinghua and San Diego datasets. For all data lengths, the proposed HTCCA method performed better in both detection accuracy and ITR. Paired t-test revealed that there were significant differences in classification accuracy between these two methods on both Tsinghua dataset and San Diego dataset ($p<0.001$ for all data length on both datasets).

## IV. DISCUSSION

The performance comparison of the three methods showed the effectiveness of the HTCCA method. When the training data are limited, the proposed HTCCA method significantly outperformed the other two methods.

Theoretically, if there are enough training data from the specific subject, the proposed HTCCA method should be inferior to TDCCA, because in this case, individual difference between subjects may have more negative influence. Transfer learning based on the training data of independent subjects is beneficial to SSVEP detection performance only when the training data of specific subject are not enough, and the transferred knowledge cannot replace the knowledge from the specific subject [7]. The experimental results of this study also verified the above theoretical analysis. When the number of training trials was big enough as shown in Fig.2, the results of HTCCA are inferior to those of TDCCA.

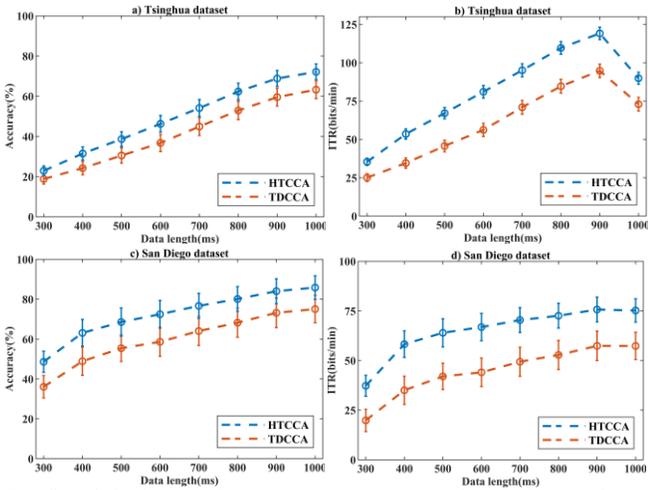

Fig. 3. Detection performance comparison between two methods when the number of trials equals to 2. (a) and (b) represent the average accuracy and ITR across subjects of Tsinghua dataset in different data lengths. (c) and (d) represent average accuracy and ITR of San Diego dataset across subjects in different data lengths. Error bars denote the standard error.

The contribution of this study is to provide a way to construct better templates by combining training information from both the specific subject and other independent subjects. Considering that in practical applications, it is time-consuming to collect enough data from a user, so it is necessary to find an effective method to collect more information from available EEG of other subjects. In addition to the significant improvement in detection performance when EEG trials are limited, this proposed HTCCA method is also easy to implement. When training data from other subjects are available, the target subject does not need to participate in lots of experiments to prepare a large amount of training data to obtain satisfactory detection performance. Therefore, the proposed HTCCA effectively avoids the problem of subject fatigue and is promising in practical use.

The Pearson correlation coefficient $\rho_1(f_i)$ in HTCCA is equivalent to $\rho(f_i)$ introduced in TDCCA, which is the key feature for SSVEP recognition. In this method, the overall accuracy depends more on the performance of key feature. This method can also be transplanted to more advanced method like TRCAR [10] to get better performance. Our future work will focus on how to combine independent data more effectively so as to improve the detection accuracy.

## V. CONCLUSION

This study proposes and evaluates the HTCCA method, which effectively improves the detection performance of SSVEP by combining the training information of specific subjects with the training information of independent subjects. The experimental results based on both Tsinghua Dataset (35 subjects) and San Diego Dataset (10 subjects) showed that the proposed method had higher performance than the existing methods in terms of detection accuracy and ITR under data limited condition. Because the user experience must be considered while achieving accurate identification accuracy, HTCCA which can achieve better performance under data-limited condition can be regarded as a promising SSVEP detection method for future BCI application.

ACKNOWLEDGMENT

This work was financed by the National Natural Science Foundation of China (Project No. 51977020). There were no experiments in this study that required the approval of the relevant committee.